\documentclass{article}  

\usepackage{amsmath,amssymb,amsfonts}
\newtheorem{theorem}{Theorem}[section]
\newtheorem{lemma}[theorem]{Lemma}

\title{\bf Spin-2 fields on Minkowski space near space-like and null
infinity.}
\author{Helmut Friedrich\\ 
Max-Planck-Institut f\"ur Gravitationsphysik\\
Am M\"uhlenberg 1\\
14476 Golm\\
Germany}

\begin{document}
\maketitle

\begin{abstract}
We show that the spin-2 equations on Minkowski space in the gauge of the
`regular finite initial value problem at space-like infinity' imply
estimates which, together with the transport equations on the cylinder at
space-like infinity, allow us to obtain for a certain class of initial
data  information on the behaviour of the solution near space-like and null
infinity of any desired precision.
\end{abstract}

\newpage

\section{Introduction}

In a recent article (\cite{chrusciel:delay}) Chru\'sciel and Delay have
shown the existence of non-trivial solutions to Einstein's vacuum
field equations which satisfy Penrose's condition of asymptotic
simplicity  (\cite{penrose:scri:let}, \cite{penrose:scri}) with a
prescribed smoothness of the asymptotic structure. The basic step in
\cite{chrusciel:delay} consists in the construction of a specific class of
asymptotically flat Cauchy data of prescribed smoothness on $\mathbb{R}^3$
which coincide with Schwarzschild data outside a fixed radius and which
can be chosen with this property arbitrarily close to Minkowski data.
Since the evolution of the data is Schwarzschild near space-like
infinity, one has perfect control on the asymptotic structure near
space-like infinity and one can construct hyperboloidal data arbitrarily
close to Minkowskian hyperboloidal data. A general result of
\cite{friedrich:n-geod} on the hyperboloidal initial value problem then
implies the existence of the desired space-times. 

While this result finally settles a question which raised some controversy
and remained open for forty years, the technique used in 
\cite{chrusciel:delay} precludes the possibility to resolve the main open
question concerning asymptotic simplicity: how `large' is the class of
asymptotically simple solutions, or, more precisely, how can this class be
characterized in terms of Cauchy data ? The answers to this
question requires a general and very detailed analysis of the behaviour of
the solutions in the domain where `null infinity touches space-like
infinity' and the data constructed in \cite{chrusciel:delay} are designed
precisely to circumvent this task.

A basic step towards answering this question has been taken in 
\cite{friedrich:i-null}, using the general conformal representation of
Einstein's vacuum field equations introduced in \cite{friedrich:AdS}. This
allows us to employ a gauge based on conformally invariant structures, which
simplifies the analysis of the equations and of the underlying conformal
geometry. It allows us in particular to control, in the given gauge, the
location of null infinity in terms of the initial data. Under suitable
assumptions on the initial data it has been shown that the standard Cauchy
problem can be reformulated to obtain a `regular finite initial value
problem near space-like infinity'.  Since this requires the complete
information on the asymptotic structure of the initial data near space-like
infinity, the data have been assumed in \cite{friedrich:i-null} to be
time-symmetric. Recent results on non-time-symmetric data
\cite{dain:friedrich} will make it possible to extend the analysis of
\cite{friedrich:i-null} to more general space-times.

In the regular finite initial value problem near space-like infinity, the
initial hypersurface $S$ is a three-manifold diffeomorphic to a closed ball
in $\mathbb{R}^3$ (possibly with its center removed). Its spherical
boundary, denoted by $I^0$, represents space-like infinity for the initial
data on
$S$.  With respect to the solution space-time space-like infinity is
represented by a cylinder $I$ diffeomorphic to
$\,]-1 , 1[ \times I^0$, which intersects $S$ at $I^0$. 

There are coordinates $\rho \ge 0$ with $\rho = 0$ on $I$ and $\tau$ with
$\tau = 0$ on $S$. The range of $\tau$ is limited by continuous functions
$\tau_{\pm}(p)$, $p \in S$, with $\tau_- \le -1$, $\tau_+ \ge 1$ and
$\tau_{\pm}(p) \rightarrow \pm 1$ as $p \rightarrow I^0$, such that the
sets ${\cal J}^{\pm} = \{\tau = \tau_{\pm}, \rho > 0\}$ represent
future and past null infinity. These sets `touch' the cylinder $I$ at the
two components $I^{\pm} = \{\tau = \pm 1, \rho = 0\}$ of its boundary,
which are diffeomorphic to $I^0$.

While the reduced equations are symmetric hyperbolic on the `physical' part
of the underlying manifold, where $\rho > 0$, and extend with this property
to the cylinder $I$, they develop a degeneracy at the spherical sets
$I^{\pm}$.  It turns out that this degeneracy is not a deficiency of our  
formulation. As argued in the following it rather indicates in a precise
way certain innate features of the generic asymptotic structure.
Understanding the consequences of this degeneracy is the main open
problem of the subject of asymptotics. Its resolution will provide a host
of detailed information on asymptotically flat space-times, new insight
into the field equations, and it will open the door for various
theoretical and practical applications.

The cylinder $I$ represents a boundary of the physical manifold. However, it
is not a boundary for the conformal field equations in the sense that one
could prescribe boundary data on $I$. The evolution of the unknown $u$ in
the conformal field equations, which comprises frame and connection
coefficients and the components of the non-physical Ricci tensor and
the rescaled conformal Weyl tensor, is governed on $I$ by an intrinsic
system of {\it transport equations} induced
on $I$ by the conformal field equations. In fact, 
for suitably chosen smooth initial data, the complete system of derivatives
$u^p = \partial_{\rho}^p u|_I$, $p = 0, 1, 2, \ldots$ , is determined
by intrinsic symmetric hyperbolic transport equations on $I$ and
corresponding initial data on $I^0$.

It has been shown in \cite{friedrich:i-null} that the transport equations
can be solved explixcitly to arbitrary order, provided certain algebraic
complexities can be handled (e.g. with an algebraic computer program). 
In \cite{friedrich:kannar1} it has been demonstrated that our analysis
can be related to earlier analyses of null infinity and that the functions
$u^p$ supply interesting information on the space-time near space-like and
null infinity. We expect that under suitable assumptions on the initial
data the setting developed in \cite{friedrich:i-null} will allow us to
obtain perfect control on the asymptotic structure of space-time.

The explicit calculation of certain components of the functions $u^p$
on $I$ shows that, in general, logarithmic singularities of the
form $(1 - \tau)^k\,\log^j(1 -\tau)$ will occur in the $u^p$ near $I^+$,
where $\tau \rightarrow 1$. These singularities can even be observed for
data which are analytic at space-like infinity. It turns out, however, that
they vanish for data which satisfy a certain set of `regularity conditions'
(cf. \cite{friedrich:i-null} for details and \cite{friedrich:tueb} for a
discussion of the related conjecture concerning the solution space-time and
an associated subconjecture concerning the smoothness of the functions
$u^p$ on $\bar{I} = I \cup I^+ \cup I^-$).

Due to the hyperbolicity of the reduced equations, the tendency
of the quantities $u^p$ to become singular at $I^{\pm}$ is likely to
spread along characteristics sufficiently close to null infinity, thus
destroying the possibility to define an asymptotic structure at
null infinity of the desired smoothness if the regularity conditions are
not satisfied to a sufficiently high order. But even if the latter are
satisfied at all orders and if it can be shown that they imply the
smoothness of the functions $u^p$ on $\bar{I}$, the decision about the
existence of a smooth extension of the solution to null infinity  is
still complicated because the degeneracy of the equations at $I^{\pm}$
interferes with the known techniques to deduce estimates for solutions
of symmetric hyperbolic systems. 

This requires us to go beyond the general theory and to look for specific
features of the conformal field equations which might allow us to overcome
this difficulty.  

The degeneracy occurs in the part of the reduced equations
which is deduced from the Bianchi equation 
$\nabla_{\mu}\,d^{\mu}\,_{\nu \lambda \rho} = 0$ for the rescaled conformal
Weyl tensor $d^{\mu}\,_{\nu \lambda \rho} = \Theta^{-1}\,C^{\mu}\,_{\nu
\lambda \rho}$. All the other equations remain regular near $I^{\pm}$. The
type of degeneracy remains if the equations are linearized at Minkowski
space in the specific gauge near space-like infinity introduced in
\cite{friedrich:i-null}.  We are thus led to analyse the spin-2 equation
on Minkowski space in this particular gauge. It is the purpose of the
present article to discuss this situation. We shall show that in the
given setting the spin-2 equations imply certain estimates, which,
combined with the transport equations on the cylinder at space-like
infinity, yield any desired information on the asymptotic behaviour of
the solution (cf. the relations (\ref{asexp}), (\ref{assmooth})). This
result on the solution will not be stated as a theorem (cf. also the
remarks following (\ref{assmooth})), because our main interest here lies
in the nature of argument and in the underlying particularities of the
setting and the equations. We expect that arguments of this type will
also help us tackle the quasi-linear problem.

\section{Minkowski space near space-like and null infinity}

Let $y^{\mu}$ be coordinates on Minkowski space in which the metric takes
is standard form $\tilde{g} = \eta_{\mu\nu}\,dy^{\mu}\,dy^{\nu}$ with 
$\eta_{\mu\nu} = diag(1,-1,-1,-1)$. We want to prescibe initial data for
the spin-2 equation on the hypersurface $\tilde{S} = \{y^0 = 0\}$ and
study the behaviour of the solution on the conformal extension
of Minkowski space  in a neighbourhood of space-like infinity which
includes a part of null infinity.  

To discuss the field on such a neighbourhood, we perform on 
$N =  \{y_{\mu}\,y^{\mu} < 0\}$ the coordinate transformation
$y^{\mu} \rightarrow x^{\mu} = - \frac{y^{\mu}}{y_{\nu} y^{\nu}}$ and
formally extend the domain of validity of the coordinates $x^{\mu}$ to
include the part ${\cal J}'$ of the set $\{x_{\mu}\,x^{\mu} = 0\}$ adjacent
to $N$. In the new coordinates, or in the associated spatial polar
coordinates with radial coordinate $\rho = \sqrt{\sum_{\mu = 1}^3
(x^{\mu})^2}$, the metric takes the forms 
\[
\tilde{g} = \frac{1}{(x_{\lambda} x^{\lambda})^2}\,
\eta_{\mu\nu}\,dx^{\mu}\,dx^{\nu} =
\frac{1}{(\rho^2 - (x^0)^2)^2}\,
((dx^0)^2 - d \rho^2 - \rho^2\,d \sigma^2),
\] 
where $d\sigma^2$ denotes the standard line element on the 2-sphere. 
Introducing the conformal factor $\Omega = - x_{\lambda} x^{\lambda}$
on $N$ we find that $\Omega$ and $g' \equiv \Omega^2\,\tilde{g}$ extend
smoothly to the set ${\cal J}'$. The conformal factor $\Omega$ vanishes
there but $g'$ remains regular. The point $i^0 = \{x^{\mu} = 0\}$ then
represents spacelike infinity and the sets
${\cal J}^{\pm} = {\cal J}' \cap \{x_{\mu}\,x^{\mu} = 0, \,\pm x^0 > 0\}$
represent parts of future resp. past null infinity  of Minkowski space.

We can reconstruct this representation of Minkowski space by solving the
conformal field equations with data on the set $S' = \{x^0 = 0\}$ which
are given by the intrinsic 3-metric $h = - (d \rho^2 + \rho^2\,d \sigma^2)$
and the second fundamental form $\chi = 0$ induced by $g'$, by the
conformal factor $\Omega = \rho^2$, and by certain fields derived from
$h$ and $\Omega$. If we slightly perturb $h$ and $\chi$ now (keeping
the conformal constraints satisfied) to obtain more general solutions, the
rescaled conformal Weyl tensor will develop a singularity at the point 
$i = \{x^{\mu} = 0\}$ in $S'$ as soon as the data acquire a non-vanishing
ADM-mass. The following gauge arose from the desire to analyse this
situation (cf. \cite{friedrich:i-null}).

To define a different scaling of the metric and a new coordinate $\tau$
we choose a function $\kappa = \rho\,\mu$, where $\mu$ is a smooth
positive function of $\rho \in \mathbb{R}$ satisfying $\mu(0) = 1$, and
set $x^0 = \tau\,\kappa$. With the conformal factor 
\begin{equation}
\label{confac}
\Theta = \frac{\rho}{\mu}\,(1 - \tau^2\,\mu^2)
= \frac{1}{\kappa}\,\Omega,
\end{equation} 
we then find
\begin{equation}
\label{metric}
g \equiv \Theta^2\,\tilde{g} = \frac{1}{\kappa^2}\,
\left(\kappa^2\,d\tau^2 + 2\,\tau\,\kappa\,\kappa'\,d\tau\,d\rho
- (1 - \tau^2\,\kappa^{'2})\,d\rho^2 - \rho^2\,d\,\sigma^2\right).
\end{equation}
With $\rho$ and $\tau$ and suitable spherical coordinates we have 
$N = \{\rho > 0,\, - \frac{1}{\mu(\rho)} < \tau < \frac{1}{\mu(\rho)}\}$
and ${\cal J}^{\pm} = \{\rho > 0,\, \tau = \pm \frac{1}{\mu(\rho)}\}$. We
set $I = \{\rho = 0,\, |\tau| < 1\}$, $I^0 = \{\rho = 0,\, \tau = 0\}$, 
and denote by $I^{\pm} = \{\rho = 0,\,\,\tau = \pm 1\}$ the sets where $I$
`touches' ${\cal J}^{\pm}$. It is understood here that $I$ is diffeomorphic
to $[-1 , 1] \times S^2$, $I^0$ and $I^{\pm}$ are diffeomorphic to $S^2$
and that the spherical coordinates extend as smooth coordinates to $I$ and
$I^{\pm}$. Finally, we set $ S = \tilde{S} \cup I^0$,
$\bar{N} = N \cup {\cal J}^+ \cup {\cal J}^- \cup \bar{I}$ with
$\bar{I} = I \cup I^+ \cup I^-$ and consider $\rho$, $\tau$ and the
spherical coordinates as coordinates on $\bar{N} \simeq 
[-1 ,1] \times [0, \infty[ \times S^2$. In these
coordinates the expressions above for $\Theta$ and $\Omega = \kappa\,\Theta$
extend smoothly to $\bar{N}$ and the coordinate expression for $g$ extends
smoothly and without degeneracy to $N \cup {\cal J}^+ \cup {\cal J}^-$
while it becomes singular on $\bar{I}$. In this new representation
the set $I^0$ corresponds to $i$ and with respect to $N$ space-like
infinity is represented now by the cylinder $I$, which can be regarded as
a kind of `blow-up' of the point $i^0$. Note that the differential
structure defined here near $I$ is completely different from the
differential structure defined by the coordinates $x^{\mu}$ near
$i^0$.  

To write out the spin-2 equation $\nabla^f\,_{a'}\,\phi_{abcf} = 0$ we
introduce a pseudo-orthonormal frame $c_{aa'}$ satisfying 
$g(c_{aa'}, c_{bb'}) = \epsilon_{ab}\,\epsilon_{a'b'}$ and
$\bar{c}_{aa'} = c_{aa'}$. As real null
vector fields we choose 
\[
c_{00'} = \frac{1}{\sqrt{2}}\left\{
(1 - \kappa'\,\tau)\,\partial_{\tau}
+ \kappa\,\partial_{\rho}\right\},\,\,\,\,\,\, 
c_{11'} = \frac{1}{\sqrt{2}}\left\{
(1 + \kappa'\,\tau)\,\partial_{\tau}
- \kappa\partial_{\rho}\right\},
\]
on $\bar{N} \setminus \bar{I}$. We note that with these conventions the
(linearized) radiation field on ${\cal J}^+$ and the null data on the
outgoing null hypersrufaces tangent to $c_{11'}$ correspond to the
components $\phi_0 \equiv \phi_{0000}$ and $\phi_4 \equiv \phi_{1111}$
respectively.

The vectors $c_{01'}$, $c_{01'}$ are then
necessarily tangent to the spheres $\tau, \rho = const.$ and can not define
smooth global vector fields. Since this leads to various arkward
expressions, all fields $c_{01'}$, $c_{01'}$ satisfying the normalization
condition above will be considered. We thus define a 5-dimensional
subbundle of the bundle of frames with structure group $U(1)$ which
projects onto $\bar{N} \setminus \bar{I}$, the projection corresponding to
the standard Hopf map $SU(2) \rightarrow SU(2)/U(1) \simeq S^2$. We lift
all our structures on $\bar{N} \setminus \bar{I}$ to this subbundle and
keep the notation used before. 

Allowing $\rho$ to take the value $0$, we extend everything, including the
projection. We consider thus $\tau$, $\rho$, and $s \in SU(2)$ as
`coordinates' on the extended subbundle, which we denote again by
$\bar{N}$. The lifted conformal factor and the lifted metric are again
given by (\ref{confac}) and (\ref{metric}), with
$d\,\sigma^2$ denoting the pull back of the line element on $S^2$ to
$SU(2)$. Thus $\bar{N}$ is difffeomorphic to $]-1, 1[ \times [0, \infty[
\times SU(2)$ now and the sets $I$, $I^{\pm}$, ${\cal J}^{\pm}$ etc. are now
considered as subsets of $\bar{N}$ which are defined by the same conditions
on $\tau$ and $\rho$ as before.

We consider the basis
\begin{equation}
\label{generators}
u_1 = \frac{1}{2} \left( \begin{array}{cc}
0 & i \\
i & 0
\end{array} \right), \, \, 
u_2 = \frac{1}{2} \left( \begin{array}{cc}
0 & - 1 \\
1 & 0
\end{array} \right),\,\,
u_3 = \frac{1}{2} \left( \begin{array}{cc}
i & 0 \\
0 & - i
\end{array} \right),
\end{equation}
of the Lie algebra of $SU(2)$ with commutation relations 
$[u_i, u_j] = \epsilon_{ijk}\,u_k$ and denote by $Z_i$, $i = 1,2, 3$, the
(real) left invariant vector field generated by $u_i$ on the (real) Lie
group $SU(2)$. We consider $Z_3$ as the vertical vector field which
generates the group $U(1)$ acting on the fibres of $\bar{N}$
and set $X = - 2\,i\,Z_3$. Defining the complex conjugate vector fields
$X_{\pm} = - (Z_2 \pm i\,Z_1)$ and setting
\[
c_{01'} = - \frac{1}{\sqrt{2}}\,\mu\,X_+,\,\,\,\,\,\, 
c_{01'} = - \frac{1}{\sqrt{2}}\,\mu\,X_-,
\]
we obtain smooth vector fields $c_{aa'}$, $Z_3$ on 
$\bar{N} \setminus \bar{I}$, which extend smoothly to $\bar{I}$ and satisfy
$g(Z_3,\,.\,) = 0$, and $g(c_{aa'}, c_{bb'}) =
\epsilon_{ab}\,\epsilon_{a'b'}$ on $\bar{N} \setminus \bar{I}$.

The connection form induced on $\bar{N} \setminus \bar{I}$ defines
connection coefficients with respect to $c_{aa'}$, which take the values
\[
\Gamma_{00'01}  = 
\Gamma_{11'01}  = - \frac{1}{2\,\sqrt{2}}(\mu + \rho\,\mu'),\,\,\,\,
\,\,\,\,
\Gamma_{01'11}  =
\Gamma_{10'00}  = \frac{1}{\sqrt{2}}\rho\,\mu',
\]
\[
\Gamma_{00'00}  = 
\Gamma_{00'11}  = 
\Gamma_{01'00}  = 
\Gamma_{01'01}  = 
\Gamma_{10'01}  = 
\Gamma_{10'11}  = 
\Gamma_{11'00}  = 
\Gamma_{11'11}  = 0,
\]
and extend smoothly to $\bar{I}$. 

Making use of the conformal covariance of the spin-2 equations, we write
them in the new conformal gauge in terms of the frame and the connection
coefficients above and obtain 
\begin{equation}
\label{kconfminkb}
0 = (1 + (\mu + \rho\,\mu')\tau)\,\partial_{\tau}\phi_k
- \rho\,\mu\partial_{\rho}\phi_k + \mu\,X_+\,\phi_{k+1}
+ ((2 - k)\,\mu + 3\,\rho\,\mu')\,\phi_k,
\end{equation}
\begin{equation}
\label{k+1confminkb}
0 = (1 - (\mu + \rho\,\mu')\tau)\,\partial_{\tau}\phi_{k+1}
+ \rho\,\mu\partial_{\rho}\phi_{k+1} + \mu\,X_-\,\phi_k
+ ((1 - k)\,\mu - 3\,\rho\,\mu')\,\phi_{k+1},
\end{equation}
for $k = 0, 1, 2, 3$, where we set as usual $\phi_0 = \phi_{0000}$,
$\phi_1 = \phi_{1000}$, etc. Notice that the coefficients of these equations
are defined and smooth for all values of $\tau$, $\rho$ and $s$ and the
equations thus make sense and are regular on $\bar{N}$.

These equations can be obtained immediately from equations given in 
\cite{friedrich:i-null} by linearizing the latter, in the given gauge, at
Minkowski space. The underlying construction, which may appear rather
arbitrary here, has a geometrical background for which we refer the
reader to
\cite{friedrich:i-null}. In particular, the curves on which $\rho$
and $s$ are constant are conformal geodesics with parameter $\tau$, the set
$I$ is attained as a limit set of these curves, and the field $c_{aa'}$ are
parallely propagated along these curves with respect to a certain Weyl
connection for $g$.

\section{The spin-2 equations near space-like and null infinity}

Equations (\ref{kconfminkb}), (\ref{k+1confminkb}) imply the
equivalent system of evolution equations 
\[
0 = (1 + \kappa'\,\tau)\,\partial_{\tau}\,\phi_0 
- \kappa\,\partial_{\rho}\,\phi_0 + \mu\,X_+\,\phi_1
+ (2\,\mu + 3\,\rho\,\mu')\,\phi_0,
\]
\[
0 = 2\,\partial_{\tau}\,\phi_1
+ \mu\,X_+\,\phi_2
+ \mu\,X_-\,\phi_0
+ 2\,\mu\,\phi_1,
\]
\[
0 = 2\,\partial_{\tau}\,\phi_2
+ \mu\,X_+\,\phi_3
+ \mu\,X_-\,\phi_1
\]
\[
0 = 2\,\partial_{\tau}\,\phi_3
+ \mu\,X_+\,\phi_4
+ \mu\,X_-\,\phi_2
- 2\,\mu\,\phi_3,
\]
\[
0 = (1 - \kappa'\,\tau)\,\partial_{\tau}\,\phi_4 
+ \kappa\,\partial_{\rho}\,\phi_4 + \mu\,X_-\,\phi_3
- (2\,\mu + 3\,\rho\,\mu')\,\phi_4,
\]
and the constraints 
\[
0 = \kappa'\,\tau\,\partial_{\tau}\phi_1  - \kappa\,\partial_{\rho}\phi_1
+ \frac{\mu}{2}\,(X_+ \phi_2 - X_- \phi_0) + 3\,\rho\,\mu'\,\phi_1,
\]
\[
0 = \kappa'\,\tau\,\partial_{\tau}\phi_2  - \kappa\,\partial_{\rho}\phi_2
+ \frac{\mu}{2}\,(X_+ \phi_3 - X_- \phi_1) + 3\,\rho\,\mu'\,\phi_2,
\]
\[
0 = \kappa'\,\tau\,\partial_{\tau}\phi_3  - \kappa\,\partial_{\rho}\phi_3
+ \frac{\mu}{2}\,(X_+ \phi_4 - X_- \phi_2) + 3\,\rho\,\mu'\,\phi_3.
\]
The latter reduce to interior equations and thus imply conditions on the
data on $S = \{\tau = 0\} \subset \bar{N}$. We shall not study them in any
detail here.

For the following discussion it is convenient to assume that $\mu' < 0$ for
$\rho > 0$. The evolution equations are then symmetric hyperbolic on
the set $\bar{N} \setminus (I^+ \cup I^-)$. Writing them in the form
$A^{\mu}\,\partial_{\mu}\,\phi = B\,\phi$, where $\phi$ denotes a column
vector with entries $\phi_k$, we get  
$A^{\tau} = diag(1 + \kappa'\,\tau,\, 2,\, 2,\, 2,\, 1 - \kappa'\,\tau)$.
Thus, while being positive definite on $\bar{N} \setminus (I^+ \cup I^-)$
and ensuring by this the hyperbolicity of the evolution equations, the
matrix $A^{\tau}$ looses this property on $I^{\pm}$ and it does so for any
choice of $\mu$ as above. It will be seen below that this renders the
standard energy estimates useless at $I^{\pm}$. This
degeneracy represents the central problem of our discussion.

Data for the linear spin-2 equations which are smooth on the hypersurface 
$\{y^0 = 0\}$ of Minkowski space are known to develop into a smooth solution
on Minkowski space and thus imply smooth fields $\phi_k$ on $N$ in the
gauge above. Whether these fields extend smoothly to $I$ and 
${\cal J}^{\pm}$ clearly depends on the behaviour of the data near
space-like infinity. Linearizing the data considered in
\cite{friedrich:i-null} at Minkowski space, we obtain a class of non-trivial
data $\phi_k$ which extend in our gauge smoothly to all of $S$. The data in
\cite{friedrich:i-null} were chosen to be time-symmetric. Linearizing 
similarly the data for the non-linear equations obtained from those
discussed in \cite{dain:friedrich} provides a large class of
non-time-symmetric data for the linear spin-2 equation which also extend
smoothly to $I^0$.

The equations above extend smoothly across $I$ into a domain where 
$\rho < 0$ and remain symmetric hyperbolic there. Extending the data on $S$
smoothly into a region of $\{\tau = 0\}$ in which $\rho < 0$, not
necessarily observing the constraints there, and evolving the data with the
extended equations, we find that the solution on $N$ extends smoothly to
$I$. Because of the degeneracy of $A^{\tau}$ the standard theory for
symmetric hyperbolic systems does not allow us to derive statements about
the behaviour of the solutions at $I^{\pm}$ and, as a consequence, also
not on ${\cal J}^{\pm}$.

A detailed inspection of the equations and the data shows, that the
degeneracy at $I^{\pm}$ can have consequences for the asymptotic
smoothness of the solutions. It is an important feature of the evolution
equations above that the matrix
$A^{\rho}$ vanishes on $I$ and the system thus implies an intrinsic system
of transport equations on $I$ which determines the solution on $I$ in terms
of the data implied by $\phi$ on $I^0$. In fact, applying formally the
operators $\partial^p_{\rho}$ to the equations and restrincting to the set
$I$ we find that the functions $\phi^p = \partial^p_{\rho} \phi|_I$ satisfy
a system of linear symmetric hyperbolic transport equations for 
$p = 0, 1, 2, \ldots$  By expanding the functions $\phi^p$ in a suitable
function system on $SU(2)$ the equations can be reduced to systems of ODE's
which can be solved explicitely. The solutions can be read off directly
from the results in \cite{friedrich:i-null}. The details of this will not
be important for the following discussion. What is important, though,
is the fact that one obtains solutions which are regular (in fact,
polynomial) in $\tau$ but besides those for $p \ge 2$ also
solutions which are singular at $I^{\pm}$. The latter are of the form
\[
\left(\frac{1 - \tau}{2}\right)^{p - k + 2}\,
\left(\frac{1 + \tau}{2}\right)^{p + k - 2}\,I^p_k(\tau)
\]
where 
\[
I^p_k(\tau) \equiv
\int\limits_0^\tau \frac{d\,\sigma}{(1 - \sigma)^{p - k + 3}  
(1 + \sigma)^{p + k - 1}} 
\]
with constants of integration $e_k$, $f_k$ which are determined by the
data on $S$. Expanding the integral one finds that these solutions
behave like
\[
(1 - \tau)^{p - k + 2}\,
(1 + \tau)^{p + k - 2}\,\log (1 - \tau) + \,\,analytic
\]
as $\tau \rightarrow 1$. Note that with increasing $p$ the singularity
gets less severe.   
The solutions thus develop in general logarithmic singularities at
$I^{\pm}$. These singularities can be expected to spread along the
characteristics ${\cal J}^{\pm}$ of the evolution equations. 

Obviously, the occurrence of these singularities depends on the constants
$e_k$, $f_k$. In \cite{friedrich:i-null} certain `regularity conditions' on
the time symmetric data have been derived. In terms of the free data given
there, i.e. a conformal metric $h$ on $\tilde{S} = \mathbb{R}^3$ which is
assumed to extend smoothly if the latter is suitably compactified in the
form $\tilde{S} \rightarrow S' = \tilde{S} \cup \{i\} \simeq S^3$ by adding
a point $i$ at space-like infinity, these conditions read  
\begin{equation}
\label{regcond}
D_{(a_q b_q} \ldots D_{a_1 b_1}\,b_{abcd)}(i) = 0,
\,\,\,\,\,q = 0,1,2, \ldots, q_*,  
\end{equation}
where we employ the space spinor notation and $b_{abcd}$ denotes the
Cotton spinor of $h$. We expect that for given $p_*$ condition
(\ref{regcond}) ensures the non-existence of logarithmic singularities in
the $u^p$ for $p = 0, 1, \ldots, p_*$ if it is satisfied for sufficiently
large $q_*$. From the calculation in \cite{friedrich:i-null} it follows 
that this statement is true in the linearized setting considered above, if
(\ref{regcond}) is replaced by its linarization at the Minkowski data
$h = - (d \rho^2 + \rho^2\,d \sigma^2)$, $\chi = 0$. In this case the
$\phi^p$ extend smoothly, in fact as polynomials in $\tau$, to $I^{\pm}$.

However, the degeneracy of $A^{\tau}$ there still does not allows us to draw
conclusions about the smoothness of the solutions on ${\cal J}^{\pm}$ by
applying standard techniques for symmetric hyperbolic systems.

\vspace{.5cm}

For the following discussion it will be convenient to use a more specific
gauge. We set $\mu \equiv 1$ such that ${\cal J}^{\pm} = \{\tau =
\pm 1,\,\rho > 0\}$.  The evolution equations then take the form
\begin{equation} 
\label{0evol}
0 = E_0 \equiv (1 + \tau)\,\partial_{\tau}\,\phi_0 -
\rho\,\partial_{\rho}\,\phi_0   + X_+\,\phi_1 + 2\,\phi_0,   
\end{equation}
\begin{equation}
\label{1evol}
0 = E_1 \equiv 2\,\partial_{\tau}\,\phi_1
+ X_-\,\phi_0 + X_+\,\phi_2 + 2\,\phi_1,   
\end{equation}
\begin{equation}
\label{2evol}
0 = E_2 \equiv 2\,\partial_{\tau}\,\phi_2
+ X_-\,\phi_1 + X_+\phi_3,
\end{equation}
\begin{equation}
\label{3evol}
0 = E_3 \equiv 2\,\partial_{\tau}\,\phi_3
+ X_-\,\phi_2 + X_+\phi_4 - 2\,\phi_3,
\end{equation}
\begin{equation}
\label{4evol}
0 = E_4 \equiv (1 - \tau)\,\partial_{\tau}\,\phi_4 
+ \rho\,\partial_{\rho}\,\phi_4  
+ X_-\,\phi_3 - 2\,\phi_4, 
\end{equation} 
and the constraints read  
 \begin{equation}
\label{1con}
0 = C_1 \equiv \tau\,\partial_{\tau}\,\phi_1 -
\rho\,\partial_{\rho}\,\phi_1 - \frac{1}{2}\,X_-\,\phi_0 +
\frac{1}{2}\,X_+\,\phi_2, 
\end{equation} 
\begin{equation}
\label{2con}
0 = C_2 \equiv \tau\,\partial_{\tau}\,\phi_2 -
\rho\,\partial_{\rho}\,\phi_2 - \frac{1}{2}\,X_-\,\phi_1 +
\frac{1}{2}\,X_+\,\phi_3, 
\end{equation} 
\begin{equation}
\label{3con}
0 = C_2 \equiv \tau\,\partial_{\tau}\,\phi_3 -
\rho\,\partial_{\rho}\,\phi_3 - \frac{1}{2}\,X_-\,\phi_2 +
\frac{1}{2}\,X_+\,\phi_4. 
\end{equation} 
Equations (\ref{0evol}), (\ref{4evol}) now degenerate everywhere on the
sets $\{\tau = \pm 1,\,\rho \ge 0\}$. However, for $\rho > 0$ this happens
only  because we have chosen the coordinate $\tau$ to be constant on the
characteristics ${\cal J}^{\pm}$.

Two other classes of characteristics which will be important for us. They
are given by 
\[
C^{\rho_*}_- = \{(1 + \tau)\,\rho = \rho_*,\,\,\rho > 0,\,|\tau| < 1\}
,\,\,\,\,\,\,\,\,\,\, 
C^{\rho_*}_+ = \{(1 - \tau)\,\rho = \rho_*,\,\,\rho > 0,\,|\tau| < 1\},
\]
where $\rho_*$ is given positive number. These correspond to spherically
symmetric outgoing and ingoing null hypersurfaces in Minkowski space.  As
$\rho_* \rightarrow 0$, the sets $C^{\rho_*}_{\pm}$ approach the sets
$\bar{I} \cup {\cal J}^{\pm}$ respectively in a limit which is
non-uniform near the sets $I^{\pm}$. 

If data for (\ref{0evol}) to (\ref{4evol}) are prescribed on the subset
$S_{\rho_1, \rho_2} = \{\rho_1 < \rho < \rho_2\}$ of $S$, with 
$0 < \rho_1 < \rho_2$, the solution will in the domain where $\tau > 0$ be
determined uniquely on the open set $D^+_{\rho_1, \rho_2} \subset N$ which
is bounded below by $S$, on the left hand side by $C^{\rho_1}_+$, and on
the right hand side by $C^{\rho_2}_-$. It follows that solutions of the
extended equations for data on $S$ which are smoothly extended through
$I^0$ depend on $\bar{N}$ only on the data on $S$.

Let $\phi_k$ be a solution of (\ref{0evol}) to (\ref{4evol}) on 
$D^+_{\rho_1, \rho_2}$. Considering $E_k = E_k[\phi_j]$ as operators acting
on the $\phi_j$, a direct calculation using the commutation relations
of the fields $Z_i$ shows that $E_j[X\phi_k - 2\,(2 -k)\,\phi_k] = 0$ on
$D^+_{\rho_1, \rho_2}$, for $j, k = 0, \ldots, 4$. It follows that 
$X\phi_k = 2\,(2 -k)\,\phi_k$ hold on $D^+_{\rho_1, \rho_2}$ if these
equations are satisfied on $S_{\rho_1, \rho_2}$, i.e. the evolution
equations preserve the spin weights.  Under the same assumptions a further
direct calculation gives on $D^+_{\rho_1, \rho_2}$ the equations
\[
\partial_{\tau}\,C_1 + \frac{1}{2}X_+ C_2 + C_1 = 0,
\]
\[
\partial_{\tau}\,C_2 + \frac{1}{2}X_+ C_3 + \frac{1}{2}X_- C_1 = 0,
\]
\[
\partial_{\tau}\,C_3 + \frac{1}{2}X_- C_2 - C_3 = 0,
\]
in which the operator $\partial_{\rho}$ does not occur. This system is
symmetric hyperbolic and implies that the quantities $C_1$, $C_2$, $C_3$
vanish on $D^+_{\rho_1, \rho_2}$ if this is the case on 
$S_{\rho_1, \rho_2}$, i.e.  the evolution equations preserve the
constraints.

\section{Structure of solutions near space-like and null infinity}

We shall assume in the following that the fields $\phi_k$ represent a
smooth solution of (\ref{0evol}) to (\ref{4evol}) in $N$ which arises from
smooth data on $\tilde{S}$ which satisfy there the constraints and have
the correct spin weights. Assuming furthermore that the data extend
smoothly to $I^0$, we can assume by the arguments above that the $\phi_k$
provide in fact a smooth solution of  (\ref{0evol}) to (\ref{3con}) on the
set $\{|\tau| < 1,\,\,\rho \ge 0\} \subset \bar{N}$. The standard argument
to derive energy estimates for (\ref{0evol}) to (\ref{4evol}) proceeds as
follows. A direct calculation gives 
\begin{equation}
\label{standest}
0 = \sum_{k = 0}^4 (\bar{\phi}_k\,E_k + \phi_k\,\bar{E}_k)
\end{equation}
\[
= \partial_{\tau}\,((1+\tau)\,|\phi_0|^2 + 2\,|\phi_1|^2 + 2\,|\phi_2|^2
+ 2\,|\phi_3|^2 + (1-\tau)\,|\phi_4|^2)
+ \partial_{\rho}\,(- \rho\,|\phi_0|^2 + \rho\,|\phi_4|^2)
\]
\[
+ X_-\,(\phi_0\,\bar{\phi}_1 + \phi_1\,\bar{\phi}_2 
+ \phi_2\,\bar{\phi}_3 + \phi_3\,\bar{\phi}_4)
+ X_+\,(\phi_1\,\bar{\phi}_0 + \phi_2\,\bar{\phi}_1 
+ \phi_3\,\bar{\phi}_2 + \phi_4\,\bar{\phi}_3)
\]
\[
+ 4\,(|\phi_0|^2 + |\phi_1|^2 - |\phi_3|^2 - |\phi_4|^2).
\]

For $t \in [0, 1]$ and $\rho_* > 0$ we set 
\[
N_t = \{0 \le \tau \le t, 0 \le \rho \le \frac{\rho_* }{1 + \tau}, s \in
SU(2)\},
\]
\[
S_t = \{\tau = t, 0 \le \rho \le \frac{\rho_* }{1 + t}, s \in
SU(2)\},
\]
\[
B_t = \{0 \le \tau \le t, \rho = \frac{\rho_* }{1 + \tau}, s \in
SU(2)\},\,\,\,\,\,
I_t = \{0 \le \tau \le t, 0 = \rho, s \in SU(2)\}.
\]

Choose now $t$ with $0 \le t < 1$.
If (\ref{standest}) is integrated over $N_t$ with respect to
$d\tau\,d\rho\,d\mu$, where $d\mu$ denotes the normalized Haar measure on
$SU(2)$, the terms involving the left invariant operators $X_{\pm}$ give no
contribution (cf. the proof of lemma \ref{leibcom}) and an application of
Gauss' law gives
\[
0 = 
\int_{S_t} ((1+t)\,|\phi_0|^2 + 2\,|\phi_1|^2 + 2\,|\phi_2|^2
+ 2\,|\phi_3|^2 + (1-t)\,|\phi_4|^2)\,d\rho\,d\mu
\]
\[ 
- \int_{S_0} (|\phi_0|^2 + 2\,|\phi_1|^2 + 2\,|\phi_2|^2
+ 2\,|\phi_3|^2 + |\phi_4|^2)\,d\rho\,d\mu
\]
\[
+ \int_{I_t} (\rho\,|\phi_0|^2 - \rho\,|\phi_4|^2)\,d\tau\,d\mu
\]

\[
+ \int_{B_t} \left\{
(n_{\tau}\,((1+\tau)\,|\phi_0|^2 + 2\,|\phi_1|^2 
+ 2\,|\phi_2|^2 + 2\,|\phi_3|^2 + (1-\tau)\,|\phi_4|^2)\right.
\]
\[
\left.
+ n_{\rho}\,(- \rho\,|\phi_0|^2 + \rho\,|\phi_4|^2)
\right\}\,d\,v\,d\,\mu
\]
\[
+ 4\,\int_{N_t} (|\phi_0|^2 + |\phi_1|^2 - |\phi_3|^2 - |\phi_4|^2)\,
d\tau\,d\rho\,d\mu,
\]
where $n_{\tau} = \nu\,\rho$ and $n_{\rho} = \nu\,(1 + \tau)$, with a
suitable positive normalizing factor $\nu$, denote the components of the
conormal to $B_t$ and $d\,v\,d\,\mu$ is the volume element induced on $B_t$.
Since the $\phi_k$ are smooth on $N_t$ the integral over $I_t$ vanishes, 
the integral over $B_t$ is non-negative and we get
\begin{equation}
\label{cruest}
(1-t)\,\int_{S_t} (|\phi_0|^2 + 2\,|\phi_1|^2 + 2\,|\phi_2|^2
+ 2\,|\phi_3|^2 + |\phi_4|^2)\,d\rho\,d\mu
\end{equation}
\[ 
\le \int_{S_0} (|\phi_0|^2 + 2\,|\phi_1|^2 + 2\,|\phi_2|^2
+ 2\,|\phi_3|^2 + |\phi_4|^2)\,d\rho\,d\mu
\]
\[
4\,\int_{N_t} (|\phi_0|^2 + 2\,|\phi_1|^2 + 2\,|\phi_2|^2
+ 2\,|\phi_3|^2 + |\phi_4|^2)\,d\,\tau\,d\rho\,d\mu,
\]
which implies by the Gronwall argument   
\[
\int_{S_t} (|\phi_0|^2 + 2\,|\phi_1|^2 + 2\,|\phi_2|^2
+ 2\,|\phi_3|^2 + |\phi_4|^2)\,d\rho\,d\mu
\]
\[ 
\le (1 - t)^{-5}\,\int_{S_0} (|\phi_0|^2 + 2\,|\phi_1|^2 + 2\,|\phi_2|^2
+ 2\,|\phi_3|^2 + |\phi_4|^2)\,d\rho\,d\mu.
\]
Estimates obtained along these lines are good enough to show the existence
of solutions to (\ref{0evol}) to (\ref{4evol}) for $\tau$ in a given range
$0 \le \tau < t_*$ with a fixed $t_* < 1$, but they give little
information on the behaviour of the solutions near $\tau = 1$. The
estimate (\ref{cruest}) could perhaps be somewhat refined, however, this
would not remove the basic difficulty of this type of estimate.

\vspace{.5cm}

To derive sharp results about the smoothness of the solution near 
${\cal J}^{\pm} \cup I^{\pm}$ we shall make use of the specific properties
of the spin-2 equations in our setting such as their overdeterminedness,
the specific structure of the coefficients of the equations, and the
existence of transport equations on $I$. 

The spin-2 equations (\ref{kconfminkb}), (\ref{k+1confminkb}) take in the
present gauge the form  
\begin{equation} 
\label{a01evol}
0 = A_k \equiv (1 + \tau)\,\partial_{\tau}\,\phi_k 
- \rho\,\partial_{\rho}\,\phi_k  
+ X_+\,\phi_{k + 1} + (2 - k)\,\phi_k,
\end{equation}
\begin{equation}
\label{b01evol}
0 = B_k \equiv (1 - \tau)\,\partial_{\tau}\,\phi_{k + 1}
+ \rho\,\partial_{\rho}\,\phi_{k + 1}
+ X_-\,\phi_k + (1 - k)\,\phi_{k + 1},      
\end{equation}
where $k = 0, \ldots, 3$.   

With the operators 
$D^{q,p,\alpha} = \partial^q_{\tau}\,\partial^p_{\rho}\,Z^{\alpha}$, where
$Z^{\alpha}$ denote the operators introduced in lemma (\ref{leibcom}), the
equation  
\[
\overline{D^{q,p,\alpha}\phi_k}\,D^{q,p,\alpha}\,A_k 
+ D^{q,p,\alpha}\phi_k\,\overline{D^{q,p,\alpha}\,A_k}
\]
\[
+ \overline{D^{q,p,\alpha}\phi_{k+1}} \,{D^{q,p,\alpha}\,B_k} 
+ D^{q,p,\alpha}\phi_{k+1} \, \overline{D^{q,p,\alpha}\,B_k} = 0,
\]
can be written
\begin{equation}
\label{nequ} 
\partial_{\tau}((1 + \tau)\,|D^{q,p,\alpha}\phi_k|^2) 
+ \partial_{\tau}((1 - \tau)\,|D^{q,p,\alpha}\phi_{k + 1}|^2)
\end{equation}
\[
- \partial_{\rho}( \rho\,|D^{q,p,\alpha}\phi_k|^2)  
+ \partial_{\rho}(\rho\,|D^{q,p,\alpha}\phi_{k + 1}|^2)
\]
\[
+
Z^{\alpha}X_+(\partial^q_{\tau}\partial^p_{\rho}\bar{\phi}_k) 
Z^{\alpha}(\partial^q_{\tau}\partial^p_{\rho}\phi_{k+1}) 
+
Z^{\alpha}(\partial^q_{\tau}\partial^p_{\rho}\bar{\phi}_{k}) 
Z^{\alpha}X_+(\partial^q_{\tau}\partial^p_{\rho}\phi_{k + 1}) 
\]
\[
+
Z^{\alpha}X_-(\partial^q_{\tau}\partial^p_{\rho}\phi_k) 
Z^{\alpha}(\partial^q_{\tau}\partial^p_{\rho}\bar{\phi}_{k+1}) 
+
Z^{\alpha}(\partial^q_{\tau}\partial^p_{\rho}\phi_{k}) 
Z^{\alpha}X_-(\partial^q_{\tau}\partial^p_{\rho}\bar{\phi}_{k + 1}) 
\]
\[
- 2\,(p - q + k - 2)\,|D^{q,p,\alpha}\phi_k|^2 
+ 2\,(p - q - k + 1)\,|D^{q,p,\alpha}\phi_{k + 1}|^2 = 0.   
\]
The numerical factors in the last two terms, which result from
the specific structure of the differential operators in (\ref{a01evol}),
(\ref{b01evol}), will play a crucial role in the following. Their signs
can be suitably adjusted by the choices of $p$ and $q$. 

Integration of (\ref{nequ}) over $N_t$ with respect to
$d\tau\,d\rho\,d\mu$ gives with Gauss' law
\[
\int_{S_t}
((1 + t)\,|D^{q,p,\alpha}\phi_k|^2 
+ (1 - t)\,|D^{q,p,\alpha}\phi_{k + 1}|^2)\,d\rho\,d\mu
\]
\[
- \int_{S_0}
(|D^{q,p,\alpha}\phi_k|^2 
+ |D^{q,p,\alpha}\phi_{k + 1}|^2)\,d\rho\,d\mu
\]
\[
+ \int_{I_t} \rho\,(|D^{q,p,\alpha}\phi_k|^2  
- |D^{q,p,\alpha}\phi_{k + 1}|^2)\,d\tau\,d\mu
\]
\[
+
\int_{B_t}
\left\{(n_{\tau}(1 + \tau) - n_{\rho}\,\rho)\,|D^{q,p,\alpha}\phi_k|^2 
+ (n_{\tau}(1 - \tau) + n_{\rho}\,\rho\,)|D^{q,p,\alpha}\phi_{k +
1}|^2)\right\}\,d\,v\,d\,\mu
\]
\[
+ \int_{N_t} 
\left\{Z^{\alpha}X_+(\partial^q_{\tau}\partial^p_{\rho}\bar{\phi}_k) 
Z^{\alpha}(\partial^q_{\tau}\partial^p_{\rho}\phi_{k+1}) 
+
Z^{\alpha}(\partial^q_{\tau}\partial^p_{\rho}\bar{\phi}_{k}) 
Z^{\alpha}X_+(\partial^q_{\tau}\partial^p_{\rho}\phi_{k + 1})\right. 
\]
\[
\left. +
Z^{\alpha}X_-(\partial^q_{\tau}\partial^p_{\rho}\phi_k) 
Z^{\alpha}(\partial^q_{\tau}\partial^p_{\rho}\bar{\phi}_{k+1}) 
+
Z^{\alpha}(\partial^q_{\tau}\partial^p_{\rho}\phi_{k}) 
Z^{\alpha}X_-(\partial^q_{\tau}\partial^p_{\rho}\bar{\phi}_{k + 1})\right\}
d\tau\,d\rho\,d\mu 
\]
\[
- 2\,(p - q + k - 2)
\int_{N_t}|D^{q,p,\alpha}\phi_k|^2 d\tau\,d\rho\,d\mu 
\]
\[
+ 2\,(p - q - k + 1)
\int_{N_t}|D^{q,p,\alpha}\phi_{k + 1}|^2 d\tau\,d\rho\,d\mu
= 0.   
\]
It follows again that the integral over $B_t$ is non-negative and the
integral over $I_t$ vanishes. For given non-negative integers $m$ and $p$
summation now yields in view of lemma (\ref{leibcom})  
\[
(1 + t)\,\sum_{q'+ p' + |\alpha| \le m}
\int_{S_t}
|D^{q',p',\alpha}(\partial^p_{\rho}\phi_k)|^2d\rho\,d\mu 
\]
\[
+ (1 - t)\,\sum_{q'+ p' + |\alpha| \le m}
\int_{S_t}
|D^{q',p',\alpha}(\partial^p_{\rho}\phi_{k + 1})|^2
\,d\rho\,d\mu
\]
\[
+ 2\,\sum_{q'+ p' + |\alpha| \le m} (p' + p - q' - k + 1)\,
\int_{N_t}|D^{q',p',\alpha}(\partial^p_{\rho}\phi_{k + 1})|^2
d\tau\,d\rho\,d\mu   
\]
\[
\le \sum_{q'+ p' + |\alpha| \le m}\int_{S_0}
(|D^{q',p',\alpha}(\partial^p_{\rho}\phi_k)|^2 
+ |D^{q',p',\alpha}(\partial^p_{\rho}\phi_{k + 1})|^2)\,d\rho\,d\mu
\]
\[
+ 2\,\sum_{q'+ p' + |\alpha| \le m}(p' + p - q' + k - 2)\,
\int_{N_t}|D^{q',p',\alpha}(\partial^p_{\rho}\phi_k)|^2
d\tau\,d\rho\,d\mu.
\]
With $p > m + 2$ it follows for $k = 0, \ldots, 3$ that
\[
(1 + t)\,\sum_{q'+ p' + |\alpha| \le m}
\int_{S_t}
|D^{q',p',\alpha}(\partial^p_{\rho}\phi_k)|^2d\rho\,d\mu 
\]
\[
+ (1 - t)\,\sum_{q'+ p' + |\alpha| \le m}
\int_{S_t}
|D^{q',p',\alpha}(\partial^p_{\rho}\phi_{k + 1})|^2
\,d\rho\,d\mu
\]
\[
+ 2\,(p - m - 2)\,\sum_{q'+ p' + |\alpha| \le m} 
\int_{N_t}|D^{q',p',\alpha}(\partial^p_{\rho}\phi_{k + 1})|^2
d\tau\,d\rho\,d\mu   
\]
\[
\le 
\sum_{q'+ p' + |\alpha| \le m}\int_{S_0}
(|D^{q',p',\alpha}(\partial^p_{\rho}\phi_k)|^2 
+ |D^{q',p',\alpha}(\partial^p_{\rho}\phi_{k + 1})|^2)\,d\rho\,d\mu
\]
\[
+ 2\,( p + m + 1)\,\sum_{q'+ p' + |\alpha| \le m} 
\int_{N_t}|D^{q',p',\alpha}(\partial^p_{\rho}\phi_k)|^2
d\tau\,d\rho\,d\mu, 
\]
and thus in particular, for $k = 0, \ldots, 3$, 
\begin{equation}
\label{kle3est}
\int_{S_t} (\sum_{q'+ p' + |\alpha| \le m}
|D^{q',p',\alpha}(\partial^p_{\rho}\phi_k)|^2)\,d\rho\,d\mu 
\end{equation}
\[
\le 
\sum_{q'+ p' + |\alpha| \le m}\int_{S_0}
(|D^{q',p',\alpha}(\partial^p_{\rho}\phi_k)|^2 
+ |D^{q',p',\alpha}(\partial^p_{\rho}\phi_{k + 1})|^2)\,d\rho\,d\mu
\]
\[
+ 2\,( p + m + 1)\,
\int_{\tau = 0}^t (\int_{S_{\tau}}
(\sum_{q'+ p' + |\alpha| \le m} 
|D^{q',p',\alpha}(\partial^p_{\rho}\phi_k)|^2)\,d\rho\,d\mu)\,d\tau, 
\]
and also 
\begin{equation}
\label{crudek4est}
2\,(p - m - 2)\,\sum_{q'+ p' + |\alpha| \le m} 
\int_{N_t}|D^{q',p',\alpha}(\partial^p_{\rho}\phi_{4})|^2
d\tau\,d\rho\,d\mu   
\end{equation}
\[
\le 
\sum_{q'+ p' + |\alpha| \le m}\int_{S_0}
(|D^{q',p',\alpha}(\partial^p_{\rho}\phi_3)|^2 
+ |D^{q',p',\alpha}(\partial^p_{\rho}\phi_{4})|^2)\,d\rho\,d\mu
\]
\[
+ 2\,( p + m + 1)\,\sum_{q'+ p' + |\alpha| \le m} 
\int_{N_t}|D^{q',p',\alpha}(\partial^p_{\rho}\phi_3)|^2
d\tau\,d\rho\,d\mu. 
\]
Inequality (\ref{kle3est}) implies for $k = 0, \ldots, 3$
\[
\int_{N_t} 
(\sum_{q'+ p' + |\alpha| \le m} 
|D^{q',p',\alpha}(\partial^p_{\rho}\phi_k)|^2)\,d\tau\,d\rho\,d\mu, 
\]
\[
\le 
\frac{e^{2\,( p + m + 1)\,t} - 1}{2\,( p + m + 1)}\,
\left\{\sum_{q'+ p' + |\alpha| \le m}\int_{S_0}
(|D^{q',p',\alpha}(\partial^p_{\rho}\phi_k)|^2 
+ |D^{q',p',\alpha}(\partial^p_{\rho}\phi_{k + 1})|^2)\,d\rho\,d\mu
\right\},
\]
which gives with (\ref{crudek4est}) the estimate 
\[
\int_{N_t}
(\sum_{q'+ p' + |\alpha| \le m} 
|D^{q',p',\alpha}(\partial^p_{\rho}\phi_{4})|^2)
d\tau\,d\rho\,d\mu   
\]
\[
\le \frac{e^{2\,( p + m + 1)\,t}}{2\,(p - m - 2)}\,
\left\{\sum_{q'+ p' + |\alpha| \le m}\int_{S_0}
(|D^{q',p',\alpha}(\partial^p_{\rho}\phi_3)|^2 
+ |D^{q',p',\alpha}(\partial^p_{\rho}\phi_{4})|^2)\,d\rho\,d\mu
\right\},
\]
and thus finally, for $k = 0, \ldots , 4$ and $p > m + 2$, 

\begin{equation}
\label{maxest}
\int_{N_t} 
(\sum_{q'+ p' + |\alpha| \le m} 
|D^{q',p',\alpha}(\partial^p_{\rho}\phi_k)|^2)\,d\tau\,d\rho\,d\mu 
\end{equation}
\[
\le 
C\,\sum_{k=0}^4\int_{S_0}
(\sum_{q'+ p' + |\alpha| \le m}
|D^{q',p',\alpha}(\partial^p_{\rho}\phi_k)|^2)\,d\rho\,d\mu.
\]
The constant $C$ here depends on $p$ and
$m$ but not on $t \in [0, 1[$. Since then the right hand side does not
depend on $t$ we find that the norms on the left hand side are uniformly
bounded as $t \rightarrow 1$. Note that by using the evolution equations
(\ref{0evol}) to (\ref{4evol}) we can express the Sobolev norm on the
right hand side in terms of the initial data and their spatial
derivatives in $S$.

Observing the nature of the boundaries of the sets $N_t$ and the Sobolev
embedding theorems, we have for $0 < t \le 1$ and $ j = 0, 1, \ldots $, a
continuous embedding
\[
H^{j + 3}(int(N_t)) \rightarrow
C^{j,\lambda}(N_t),
\]
where $H$ denotes a standard $L^2$-type Sobolev space and $\lambda$
indicates a local H\"older condition of exponent $\lambda$ with $0 <
\lambda \le 1/2$. The space $C^{j,\lambda}(N_t)$ consists of functions in
$C^j(int(N_t))$ which, together with their derivatives of order $\le j$,
are locally H\"older continuous, bounded and uniformly continuous on the
interior
$int(N_t)$ of the closed set $N_t$ and thus extend together with the
derivatives of order 
$\le j$ to continuous functions on $N_t$. Observing that our 5-dimensional
setting is obtained by lifting a 4-dimensional setting we find that the
condition on $\lambda$ can be relaxed to $0 < \lambda < 1$.

By the estimate (\ref{maxest}) it follows then that for given 
non-negative integer $j$ we have
\[
\partial^p_{\rho}\,\phi_k \in C^{j,\lambda}(N_1) 
\quad\mbox{for}\quad p \ge j + 6,
\]
which allows us to get by integration the representation
\begin{equation}
\label{asexp}
\phi_k = \sum_{p = 0}^{p-1} \frac{1}{p'\,!}\,\phi^{p'}_k\,\rho^{p'}
+ J^{p}(\partial^p_{\rho}\phi_k)
\quad\mbox{on}\quad N_1
\quad\mbox{for}\quad p \ge j + 6,
\end{equation}
where $J$ denotes the operator
$f \rightarrow J(f) = \int_0^{\rho} f(\tau, r, s)\,d\,r$ and the functions 
$\phi^{p'}_k(\tau, s) = \partial^{p'}_{\rho}\,\phi_k|_I$, which are
obtained by integrating the transport equations on $I$, are
considered as being extended to $N_1$ as $\rho$-independent functions.

Since then 
\begin{equation}
\label{assmooth}
\phi_k - \sum_{p' = 0}^{p - 1} \frac{1}{p'\,!}\,\phi^{p'}_k\,\rho^{p'}
\in C^{j,\lambda}(N_1)
\quad\mbox{for}\quad p \ge j + 6,
\end{equation}
for given $j$, we can control the behaviour of the solution
near ${\cal J}^{\pm} \cup I^{\pm}$ with arbitrary precision. In
particular, if the linearization of (\ref{regcond}) is satisfies at all
orders, the functions $\phi^{p'}_k$ are smooth on $\bar{I}$ for all 
$p' = 0, 1, 2, \ldots$ and the $\phi_k$ have a smooth extension to
${\cal J}^{\pm} \cup I^{\pm}$.  

The situation cannot be expected to improve in the quasi-linear problem.
Thus the expansion above suggests that logarithmic singularities will occur
in general also in that case. This then says that we cannot find a finite
representation of the Cauchy problem near space-like infinity which is
more regular than the one obtained in \cite{friedrich:i-null}.

After introducing suitable functions spaces, the representation
(\ref{asexp}) can be used to derive estimates for the $\phi_k$ also if
logarithmic singularities are present.

We emphasize that our conclusion refers to a gauge where
${\cal J}^+ = \{\tau = 1\}$. If we had chosen $\kappa = \rho\,\mu$ with
$\mu(\rho) = 1$ for  $0 \le \rho < \rho_{**}$ but $\mu' < 0$ for $\rho >
\rho_{**}$, the simple representation (\ref{asexp}) would not be valid
on the slice $\{\tau = 1\}$ for $\rho > \rho_{**}$. Where the
coefficients in the equations will begin to differ from those of
(\ref{a01evol}), (\ref{b01evol}) the whole string of quantities
$\partial^{p'}_{\rho}\,\phi_k$, $p' < p$, may enter the estimates and the
argument needs to be modified.

In the present linear case the conclusions of (i) follows also
by a closer inspection of the quantities $\phi^p_k$ and the observation
that due to the specific nature of the equations the sum in
(\ref{assmooth}) does already define a solution (cf. \cite{valiente} for 
more details). Our point here is that we found a type of argument which is
based on features of the linearized equations which can also be
identified in the non-linear setting of \cite{friedrich:i-null}.

\section{Appendix}

The purpose of this appendix is to introduce a family of left invariant
operators on $SU(2)$ and to proof lemma (\ref{leibcom}). This implies a
considerable simplification of our estimates.

Consider the (real) left invariant vector fields $Z_i$
on
$SU(2)$. For given multi-index $\alpha = (\alpha_1, \alpha_2, \alpha_3)$
with non-negative integers $\alpha_i$ we set 
$\hat{Z}^{\alpha} = Z_1^{\alpha_1}\,Z_2^{\alpha_2}\,Z_2^{\alpha_2}$, with
the understanding that $\hat{Z}^{\alpha} = 1$ if 
$|\alpha| \equiv \alpha_1 + \alpha_2 + \alpha_3 = 0$, and consider it as a
left invariant operator on the set of smooth function on $SU(2)$.

If we identify the infinitesimal algebra over $\mathbb{R}$ generated by the
operators $Z_i$ with the universal enveloping algebra of $su(2)$,
the operators $\hat{Z}^{\alpha}$ are known to provide a basis of this
algebra. For our purposes a different basis leads to a considerable
simplification of our estimates. Writing
$\hat{Z}^{\alpha}$ in the form
\[
Z_1^{\alpha_1}\,Z_2^{\alpha_2}\,Z_2^{\alpha_2} 
= Z_{i_1} \ldots Z_{i_{\alpha_1}}\,Z_{i_{\alpha_1 + 1}}
\ldots Z_{i_{\alpha_1 + \alpha_2}}\, Z_{i_{\alpha_1 + \alpha_2 + 1}}
\ldots Z_{i_{|\alpha|}},
\]
we get from it by symmetrization and normalization the operator
\[
Z^{'\alpha} = \frac{1}{\alpha_1\,!\,\alpha_2\,!\,\alpha_3\,!}\,
\sum_{\pi \in S_{m}} Z_{\pi(i_1)} \ldots Z_{\pi(i_{|\alpha|})} = 
\sum_{A = 1}^\frac{m\,!}{\alpha_1\,!\,\alpha_2\,!\,\alpha_3\,!} 
\sigma_A,
\]
where $S_m$ denotes the symmetric group and the terms $\sigma_A$
realize the $\frac{m\,!}{\alpha_1\,!\,\alpha_2\,!\,\alpha_3\,!}$
possibilities to form out of $\alpha_i$ (indistiguishable) operators
$Z_i$, $i = 1, 2, 3$, products of $|\alpha|$ operators. For $|\alpha|
\le 1$ we have $Z^{'\alpha} = \hat{Z}^{\alpha}$.
Then equations of the form 
\[
\frac{m\,!}{\alpha_1\,!\,\alpha_2\,!\,\alpha_3\,!}\,\hat{Z}^{\alpha} =
Z^{'\alpha}  + \sum_{|\beta| < |\alpha|} a_{\beta}\,\hat{Z}^{\beta}
= Z^{'\alpha} 
+ \sum_{|\beta| < |\alpha|} c_{\beta}\,Z^{'\beta},
\]
hold with constant coefficients $a_{\beta}$ and $c_{\beta}$. The first
equation is obtained by commuting operators and observing the commutation
relations, the second equation is obtained by symmetrizing and
normalizing the lower order operators  
$\hat{Z}^{\beta}$. The operators $Z^{'\alpha}$ thus also form a basis of
the enveloping algebra of $su(2)$.

\begin{lemma}
\label{leibcom}
With the normalizing factors 
$f(\alpha) = 
c\,\sqrt{\frac{\alpha_1\,!\,\alpha_2\,!\,\alpha_3\,!}{|\alpha|\,!}}$,
where $c$ is a fixed positive constant, the operators
$Z^{\alpha} = f(\alpha)\,Z^{'\alpha}$ satisfy for any smooth
complex-valued functions $f$, $g$ on $SU(2)$ for $k = 1, 2, 3$ the
equation    
\begin{equation}
\label{grid}
\sum_{|\alpha| = m} (Z^{\alpha}\,Z_k\,f\,Z^{\alpha}\,g
+ Z^{\alpha}\,f\,Z^{\alpha}\,Z_k\,g) =
Z_k\,(\sum_{|\alpha| = m} Z^{\alpha}\,f\,Z^{\alpha}\,g).
\end{equation}
In particular, if $d\mu$ denotes the normalized Haar measure on $SU(2)$,
\begin{equation}
\label{zint}
\sum_{|\alpha| = m} 
\int_{SU(2)}(Z^{\alpha}\,X_{\pm}\,f\,Z^{\alpha}\,g
+ Z^{\alpha}\,f\,Z^{\alpha}\,X_{\pm}\,g)\,d\mu = 0.
\end{equation}
\end{lemma}


{\bf Proof}:
We consider $Z^{'\alpha}Z_1 = (\sum \sigma_A)\,Z_1$ and study what
happens if we commute $Z_1$ successively with the factors generating
the $\sigma_A$'s to obtain $Z_1 Z^{'\alpha}$. Each commutation of
$Z_1$ with a factor $Z_2$ in one of the $\sigma_A$'s generates a
transition $Z_2 \rightarrow [Z_2, Z_1] = - Z_3$ and thus a term
$- \sigma'_B$ of $- Z^{'(\alpha_1, \alpha_2 - 1, \alpha_3 + 1)}$.
The number of terms in $Z^{'(\alpha_1, \alpha_2 - 1, \alpha_3 + 1)}$
created by the complete commutation process is then
$\frac{\alpha_2\,|\alpha|\,!}{\alpha_1\,!\,\alpha_2\,!\,\alpha_3\,!}$.
Conversely, each term $- \sigma'_B$ of 
$- Z^{'(\alpha_1, \alpha_2 - 1, \alpha_3 + 1)}$ can be generated by the
commutation process from precisely $\alpha_3 + 1$ different terms in
$Z^{'\alpha}$ (there are $\alpha_3 + 1$ possibilites to replace in
$\sigma'_B$ one of the $Z_3$'s by a $Z_2$). The number  
$\frac{(\alpha_3 + 1)\,|\alpha|\,!}{\alpha_1\,!\,
(\alpha_2 - 1)\,!\,(\alpha_3 + 1)\,!}$ of terms $- \sigma'_B$ thus
obtained agrees with the number above. It follows that each term of 
$- Z^{'(\alpha_1, \alpha_2 - 1, \alpha_3 + 1)}$ is generated precisely
$\alpha_3 + 1$ times. An analogous consideration concerning the 
commutations of $Z_1$ with factors $Z_3$, which generate transitions $Z_3
\rightarrow [Z_3, Z_1] = Z_2$ and thus terms of
$Z^{'(\alpha_1, \alpha_2 + 1, \alpha_3 - 1)}$, shows that each term of 
$Z^{'(\alpha_1, \alpha_2 + 1, \alpha_3 - 1)}$ is generated precisely
$\alpha_2 + 1$ times. Similar results are obtained for the commutators
of $Z^{'\alpha}$ with $Z_2$ and $Z_3$. 

Setting now $Z^{\alpha} = g(\alpha)\,Z^{'\alpha}$ with an as yet
undetermined normalizing factor $g$, we thus get the relations
\[
Z^{\alpha}Z_1 - Z_1Z^{\alpha}
\]
\[
= 
- \frac{(\alpha_3 + 1)\,g(\alpha)}{g(\alpha_1,\alpha_2 - 1,\alpha_3 + 1)}
Z^{(\alpha_1, \alpha_2 - 1, \alpha_3 + 1)}
+ \frac{(\alpha_2 + 1)\,g(\alpha)}{g(\alpha_1,\alpha_2 + 1,\alpha_3 - 1)}
Z^{(\alpha_1, \alpha_2 + 1, \alpha_3 - 1)},
\]
\[
Z^{\alpha}Z_2 - Z_2Z^{\alpha}
\]
\[
= 
- \frac{(\alpha_1 + 1)\,g(\alpha)}{g(\alpha_1+1,\alpha_2,\alpha_3-1)}
Z^{(\alpha_1+1, \alpha_2, \alpha_3-1)}
+ \frac{(\alpha_3 + 1)\,g(\alpha)}{g(\alpha_1-1,\alpha_2,\alpha_3+1)}
Z^{(\alpha_1-1, \alpha_2, \alpha_3+1)},
\]
\[
Z^{\alpha}Z_3 - Z_3Z^{\alpha}
\]
\[
= 
- \frac{(\alpha_2 + 1)\,g(\alpha)}{g(\alpha_1-1,\alpha_2+1,\alpha_3)}
Z^{(\alpha_1-1, \alpha_2+1, \alpha_3)}
+ \frac{(\alpha_1 + 1)\,g(\alpha)}{g(\alpha_1+1,\alpha_2-1,\alpha_3)} 
Z^{(\alpha_1+1, \alpha_2-1, \alpha_3)}.
\]

With these relations it follows by a direct calculation that the 
$Z^{\alpha}$ satisfy equations (\ref{grid}) if  
\[
g(\alpha_1, \alpha_2-1, \alpha_3+1) =
\sqrt{\frac{\alpha_3+1}{\alpha_2}}\,g(\alpha)\,\,\,\,\,\alpha_2 \ge 1,
\]
\[
g(\alpha_1+1, \alpha_2, \alpha_3-1) =
\sqrt{\frac{\alpha_1+1}{\alpha_3}}\,g(\alpha)\,\,\,\,\,\alpha_3 \ge 1,
\]
\[
g(\alpha_1-1, \alpha_2+1, \alpha_3) =
\sqrt{\frac{\alpha_2+1}{\alpha_1}}\,g(\alpha)\,\,\,\,\,\alpha_1 \ge 1.
\]
The normalizing factors $f$ given in the lemma obey these rules.

If $u$ is the generator of a left invariant vector field $Z$ on $SU(2)$,
then
\[
\int_{SU(2)} Z\,f(s)\,d\mu(s) = 
\int_{SU(2)}\,\lim_{\lambda \to 0}\,\frac{1}{\lambda}(f(s) -
f(s\,\exp(\lambda\,u))\,d\mu(s) = 0,
\]
for any $C^1$ function $f$ on $SU(2)$ because the measure is right
invariant and we may commute the integration with taking the limit.
Therefore (\ref{zint}) follows immediately from (\ref{grid}).

\end{document}